\documentclass{ametsocV6.1_nolineno}

\title{Tipping points in overturning circulation mediated by ocean mixing and the configuration and magnitude of the hydrological cycle: A simple model}

\authors{Anand Gnanadesikan,\aff{a}\correspondingauthor{Anand Gnanadesikan,gnanades@jhu.edu} 
 Gianluca Fabiani,\aff{b} Jingwen Liu,\aff{a} Renske Gelderloos, \aff{a} G. Jay Brett \aff{c} ,Yannis Kevrekidis, \aff{a} Thomas Haine, \aff{a} Marie-Aude Pradal, \aff{a} Constaninos Siettos,\aff{b}
 and Jennifer Sleeman, \aff{c}}

     \affiliation{\aff{a}{Morton K. Blaustein Department of Earth and Planetary Science, Johns Hopkins University, Baltimore, MD, USA}\\
     \aff{b}{Modeling Engineering Risk and Complexity, Scuola Superiore Meridionale, and Department of Applied Mathematics, University of Naples Federico II, Naples, Italy}\\
\aff{c}{Johns Hopkins Applied Physics Lab, Laurel, MD, USA}}

\abstract{The current configuration of the ocean overturning involves upwelling predominantly in the Southern Ocean and sinking predominantly in the Atlantic basin. The reasons for this remain unclear, as both models and paleoclimatic observations suggest that sinking can sometimes occur in the Pacific. We present a six-box model of the overturning in which temperature, salinity and low-latitude pycnocline depths are allowed to vary prognostically in both the Atlantic and Pacific. The overturning is driven by temperature, winds, and mixing and modulated by the hydrological cycle. In each basin there are three possible flow regimes, depending on whether low-latitude water flowing into northern surface boxes is transformed into dense deep water, somewhat lighter intermediate water, or light water that is returned at the surface.  The resulting model combines insights from a number of previous studies and allows for nine possible global flow regimes. For the modern ocean, we find that although the interbasin atmospheric freshwater flux suppresses Pacific sinking, the equator-to-pole flux enhances it. When atmospheric temperatures are held fixed, seven possible flow regimes can be accessed by changing the amplitude and configuration of the modern hydrological cycle . North Pacific overturning can strengthen with either increases or decreases in the hydrological cycle, as well as under reversal of the interbasin freshwater flux.  Tipping-point behavior of both transient and equilibrium states is modulated by parameters such as the poorly constrained lateral diffusive mixing. If hydrological cycle amplitude is varied consistently with global temperature, northern polar amplification is necessary for the Atlantic overturning to collapse.}

\begin{document}

\maketitle

%
%
%
\statement
Currently the global overturning circulation involves conversion of light water to dense water in the North Atlantic, light water to  waters of intermediate density in the North Pacific, and dense water to intermediate density/light waters in the Southern Ocean. Many different factors have been invoked to explain this configuration, with atmospheric freshwater transport, basin geometry, lateral mixing, and Southern Ocean winds playing major roles. This paper develops a simple theory that combines previous theories, presents the intriguing idea that alternate configurations might be possible, and identifies multiple possible tipping points between these states.
%
%
%

%
\section{Introduction}

The fact that the transformation of light surface waters to dense deep waters is dominated by processes in the North Atlantic basin has profound implications for the ocean's  physical and biogeochemical structure \citep{gnanadesikan1999pycnocline,marinov2006BGCdivide} as well as for global climate. While some of the cold, dense water that rises to the surface in the Southern Ocean cools further and sinks to form the Antarctic Bottom Water, a significant fraction moves northward and is freshened and warmed as it is transformed into lighter Antarctic Intermediate (AAIW) and Subantarctic Mode Water (SAMW) \citep{lumpkin2007global}. This process is balanced by a sinking of North Atlantic Deep Water (NADW) in the North Atlantic. This meridionally asymmetric pattern is associated with a cross-equatorial heat transport \citep{trenberth2019observation} that means that a given northern latitude is usually several degrees warmer than its southern counterpart, also helping to keep the intertropical convergence zone and associated rainfall north of the equator \citep{zhang2005simulated}.

  A key driver of the north-south asymmetry in watermass transformation is that the westerly winds generate a net northward surface flow of water.  As noted by a number of authors \citep{toggweiler1993magnitude, toggweiler1995effect, gnanadesikan1999pycnocline} within the unblocked latitudes of the Southern Ocean this water cannot be supplied via a western boundary current carrying light subtropical water poleward along a continental boundary. Instead it must be upwelled from greater depths. Some fraction of this rising water is supplied by dense water flowing southward below the depth of ridges, while some is supplied by boluses of lighter low-latitude waters associated with mesoscale eddies \citep{johnson1989size,hallberg2001exploration,klinger2019ocean}. In ocean-only models \citep{fuckar2007interhemispheric,johnson2007reconciling,wolfe2011adiabatic} the structure of the zonally averaged overturning circulation has been shown to be highly dependent on whether surface fluxes can make the  Southern Ocean intermediate waters advected northward at the tip of Drake Passage lighter than NADW. If the nominal NADW becomes lighter than the nominal Antarctic Intermediate Water, then AAIW densities will not outcrop in the Northern Hemisphere. Given the low levels of diapycnal mixing observed away from the mixed layer, in such situations the net watermass transformation in the Southern Ocean must be small. If a steady-state is to be achieved throughout the ocean, the bulk of the northward flux of lighter waters in the Southern Ocean associated with northward Ekman transport must then be balanced by a southward flux of waters of similar density. This flux would be associated with some combination of mesoscale and stationary eddies \citep{johnson1989size,hallberg2001exploration}. 
 
In addition to the north-south asymmetry in the overturning, there is also an interbasin asymmetry, whereby no counterpart to the NADW is formed in the Pacific. In a seminal paper  \cite{warren1983jmr} discussed reasons for this asymmetry, arguing that two factors play an important role in producing a relatively fresh surface. First, he claimed that the North Pacific receives a greater net air-sea flux of freshwater than the North Atlantic. Second, he noted that the overturning that removes freshwater added to the North Pacific subpolar gyre is weak relative to the Atlantic. These two mechanisms, asymmetry in freshwater delivery and preferential flushing of the Atlantic, remain the two leading processes discussed in the literature today \citep{ferreira2018atlantic,johnson2019recent}.

In idealized coupled models these two processes work together to localize overturning in the Atlantic. In models with a wide basin (representing the Pacific) and a narrow basin (representing the Atlantic), dense water formation tends to occur preferentially in the narrow basin \citep{jones2016interbasin,jones2017size,youngs2020basin}.  Assuming the atmospheric freshwater transport $F_w^{basin}$ between the subtropical and subpolar gyres scales as the basin width $L_x^{nbasin}$, the subpolar gyre of a wide basin will receive more freshwater than a narrow basin, but still only has one coast along which to support an overturning circulation that can remove this freshwater. If the overturning circulation within a basin has magnitude $M_{basin}$, the salinity difference between high and low latitudes within that basin will scale as $\Delta S_{basin} \approx -(F_w^{basin}/M_{basin})  S_0$ where $S_0$ is some average salinity. Thus, given two idealized basins with equivalent initial overturning, the wider basin with a larger freshwater flux $F_w^{wide}>F_w^{narrow}$ will exhibit a larger salinity difference ($\Delta S_{wide}>\Delta S_{narrow}$). But since the impact of this gradient on density is to make polar waters lighter with respect to the tropics it will produce a weaker poleward density gradient. The weaker gradient will retard the overturning in the wider basin relative to the narrow basin, causing the wider basin to stand higher and pump tropical water into the narrow basin.  This interbasin transport then reinforces the overturning in the narrow basin. Note that the models cited above do not resolve the high topography of the Rockies and Andes (which block the eastward mid-latitude transport of water vapor from the Pacific to the Atlantic) or the east African highlands (which block westward tropical transport of water vapor from the Indian to the Atlantic).

Insofar as the  hydrological cycle is responsible for localizing of the overturning to the North Atlantic,  we would expect Pacific overturning only when conditions are cooler, as the amplitude of the hydrological cycle tends to track global temperatures. However, there are a number of lines of evidence that suggest that this picture may be incomplete. First, both idealized models and realistic coupled models can generate some overturning in the Pacific given modern conditions \citep{deBoer2010meridional,bahl2019variations}. As noted in the latter paper, the overturning is highly sensitive to the lateral diffusion coefficient $A_{Redi}$ associated with mesoscale eddies, whose spatial structure is poorly understood \citep{abernathey2022isopycnal}. Second, paleoceanographic evidence suggests that there have been times in the past when there was more deep water formation in the North Pacific \citep{rae2014deep,burls2017active,ford2022pliocene}, as there is evidence of lower levels of chemicals produced by the decomposition of organic material. This includes cold periods such as the Younger Dryas when the climate was presumably colder and freshwater transport was weaker than today, both between basins and from the subtropical to subpolar gyres. However it also includes the relatively warm Pliocene in which the freshwater transports were likely even stronger. Third, although the modern atmosphere does appear to transport less freshwater from the low latitudes to the high latitudes between 40$^\circ$S and 65$^\circ$N in the Atlantic relative to the Pacific \citep[as summarized in][]{ferreira2018atlantic}, it also deposits significant freshwater flux in the Arctic. As we will argue below, this re-enters the global ocean in the Atlantic and {\em freshens} the northern subpolar Atlantic and NADW relative to the northern subpolar Pacific and AAIW. This relative freshening can also be seen in models. For example in the preindustrial control version of a coarse resolution climate model  \citep[GFDL ESM2Mc,][]{galbraith2011esm2mc,pradal2014Redi} we find that the net freshwater flux to the North Pacific is 0.29 Sv, while the flux to the Atlantic and Arctic north of 40$^\circ$N is 0.44 Sv. Nonetheless, this model is still able to produce an overturning of 20 Sv in the Atlantic and 4 Sv in the Pacific. 

This paper seeks insight into the dynamics of how the configuration and magnitude of the hydrological cycle control the configuration and magnitude of the overturning using a relatively simple dynamical box model. Such approaches have a long history of giving insight into the dynamics and sensitivity of the overturning \citep{johnson2019recent}. For example the pioneering paper of \cite{stommel1961thermohaline} showed that the opposing effects of high latitude cooling and poleward atmospheric freshwater transport could combine to produce a bistable overturning describeable by two fold bifurcations. \citet{tziperman1994instability} showed that this mechanism could also operate in a fully coupled model, but that it only appeared for some initial conditions. \cite{huang1992multiple} showed the potential existence of multiple steady states when resolving deep and surface boxes in the Northern, low-latitude and Southern oceans.    \cite{gnanadesikan1999pycnocline} helped to explain how changes in Southern Ocean winds and eddies help determine the magnitude of the northern hemisphere overturning by controlling the transformation of dense water to light water.   \cite{johnson2007reconciling} extended the latter model to include prognostic equations for temperature and salinity and showed that, similar to the Stommel model, the balance between the hydrological and heat cycles could produce two stable states of overturning. In this case though, the stable state is found and controlled by the density difference between the North Atlantic Deep Water and Southern Ocean surface waters. In one state, transformation of dense to light water in the Southern Ocean and low-latitude pycnocline is balanced by overturning in the North with a relatively shallow low-latitude pycnocline. In the other, it is balanced by eddy fluxes of mass to the Southern Ocean associated with a deep low-latitude pycnocline.  \cite{jones2016interbasin} extended the \cite{gnanadesikan1999pycnocline} model to include a Pacific basin- explicitly looking at how Southern Ocean winds control the flow of water between the Pacific-Indian and Atlantic basins, arguing that this produced a deeper pycnocline in the Pacific relative to the Atlantic. \cite{gnanadesikan2018fourbox} extended the \cite{johnson2007reconciling} model to include lateral tracer mixing. This paper examined what happened when hydrological fluxes were adjusted to make a model with "incorrect" physics look like a model with "correct" physics.  While this process sometimes improved estimates of the stability of the overturning when the hydrological cycle amplitude was increased, as suggested by \citet{liu2017overlooked}, it did not always do so.

In this work we extend the model of \citet{gnanadesikan2018fourbox} to include a separate low-latitude Pacific-Indian ocean box and a high-latitude Pacific box. We include a low-latitude exchange term between the Indo-Pacific and Atlantic similar to that formulated previously \citep{jones2016interbasin} thus potentially distinguishing the "warm-water" and "cold-water" pathways by which light water enters the Atlantic basin. The model can also be seen as extending \cite{jones2016interbasin} by allowing for prognostic salinity and temperature and for a potential North Pacific overturning. The resulting model exhibits a phenomenologically rich interplay between freshwater fluxes and different regimes of overturning, predicting a number of new "tipping points" between states. 

\section{Model description}

\subsection{Equations}

\begin{figure}[t]
  \noindent\includegraphics[width=39pc,angle=0]{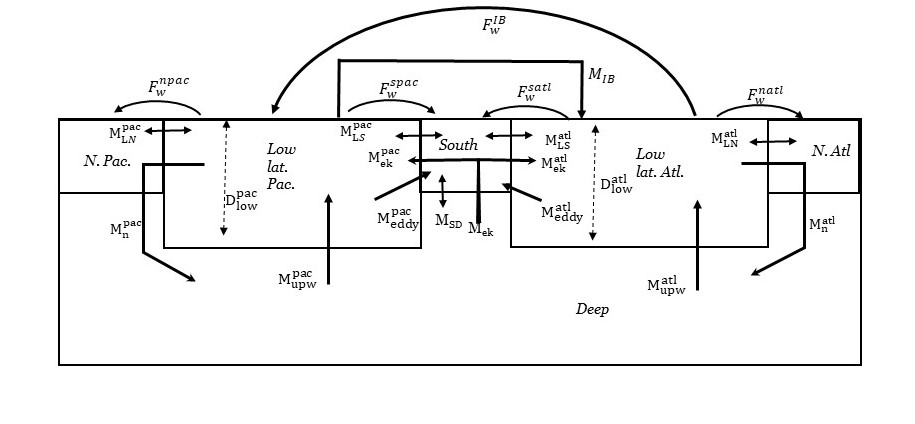}\\
  \caption{Schematic of our six box model. Note that the circulation is centered on the Southern Ocean, as in \cite{lumpkin2007global}.}
  \label{f1}
\end{figure}

A "wiring diagram" of the model is shown in Figure \ref{f1}. A full description of the model variables, parameters and equations is provided in the Supplemental Material. The model has two northern latitude boxes, each of which is capable of transforming warm, salty low-latitude water into colder, lower-salinity surface, intermediate or deep water. The high latitude boxes also exchange water and tracers with low-latitude boxes in each basin. Freshwater fluxes are specified between the low-latitude boxes and the high latitude boxes ($F_w^{npac},F_w^{natl},F_w^{spac},F_w^{satl}$) as well as between the two low-latitude boxes ($F_w^{IB}$). This allows us to examine the effect of changes in the magnitude of freshwater fluxes such as might be associated with climate change, as well as differences in the configuration of freshwater fluxes (i.e. the ratio of $F_w^{npac}$ to $F_w^{IB}$), which may vary considerably from one climate model to another. 

A key aspect of this model is that the pycnocline depth is not a fixed parameter. This contrasts with the classic box models \citep{stommel1961thermohaline, rooth1982hydrology, huang1992multiple, tziperman1994instability} previously referred to as well as the more recent versions by \cite{alkhayuon2019basin}, where the overturning throughout the ocean is directly proportional to density differences between different regions. The depths of the pycnocline in the Atlantic ($D_{low}^{atl}$) and Indo-Pacific ($D_{low}^{pac}$) are state variables whose evolution is predicted by two mass balance equations  which depend in part on density difference between the northern and southern latitudes. We define $\rho_{natl,npac,latl,lpac,S,deep}$ as the densities of the North Atlantic, North Pacific, low latitude Atlantic and low latitude Pacific/Indian, Southern Ocean surface and deep ocean respectively. Then letting $Area_{latl},Area_{lpac}$ be the area of  the low-latitude Atlantic and Pacific/Indian boxes, we can define 
\begin{align}\label{eq:vol_balance}
Area_{latl} \frac{\partial D_{low}^{atl}}{\partial t}&=M_{ek}^{atl}+M_{upw}^{atl}-M_{eddy}^{atl}-M_n^{atl} (\rho_{natl}>\rho_S)+M_{IB}-F_w^{satl}-F_w^{natl}-F_w^{IB}, \\
Area_{lpac} \frac{\partial D_{low}^{pac}}{\partial t}&=M_{ek}^{pac}+M_{upw}^{pac}-M_{eddy}^{pac}-M_n^{pac} (\rho_{npac}>\rho_S)-M_{IB}-F_w^{spac}-F_w^{npac}+F_w^{IB},
\end{align}
where the $M_{ek}$ fluxes represent the upwelling of dense water into the Southern Ocean mixed layer and its subsequent export into the mid-latitude pycnocline; the $M_{eddy}$ fluxes represent the supply of light water to the Southern Ocean driven by eddy thickness fluxes; the $M_{upw}$ terms represent diffusive upwelling in the pycnocline; and $M_{IB}$ represents the exchange of mass between the basins. Terms of the form $\rho_A>\rho_B$ are set to 1 if true and zero if false. We note that the northern overturning only changes the volume of the low-latitude box if the high-latitude box is  {\em denser than that of the Southern Hemisphere surface box}. If it is not, the assumption is that the water entering the high latitude box will return to the low-latitude box.

As in previous work \citep{gnanadesikan1999pycnocline,gnanadesikan2018fourbox} the model describes the water fluxes between the boxes in terms of parameters and state variables. Following \cite{gnanadesikan1999pycnocline} and \cite{gnanadesikan2018fourbox} the overturning fluxes in each basin $M_n^{atl}$ and $M_n^{pac}$ are taken as proportional to the depth-integrated geopotential. This can be written as 
\begin{equation}\label{eq:northern_overturn}
    M_n^{atl,pac}=\frac{g(\rho_{natl,npac}-\rho_{latl,lpac}) {D_{low}^{atl,pac}}^2}{\rho_{natl,npac} \epsilon_{natl,npac}} .
\end{equation}
Such a relationship has been found to hold in models \citep{levermann2010atlantic} and in laboratory experiments \citep{park1999rotating}.   In this equation, the $\epsilon$ terms can be thought of as the resistance to converting the depth integrated difference in geopotential to overturning. This difference is approximately proportional to the available potential energy (APE). It should be noted that these terms cover a multitude of physical processes and geometric effects, a point to which we will return later in this manuscript. 
\begin{figure}[t]
  \noindent\includegraphics[width=38pc,angle=0]{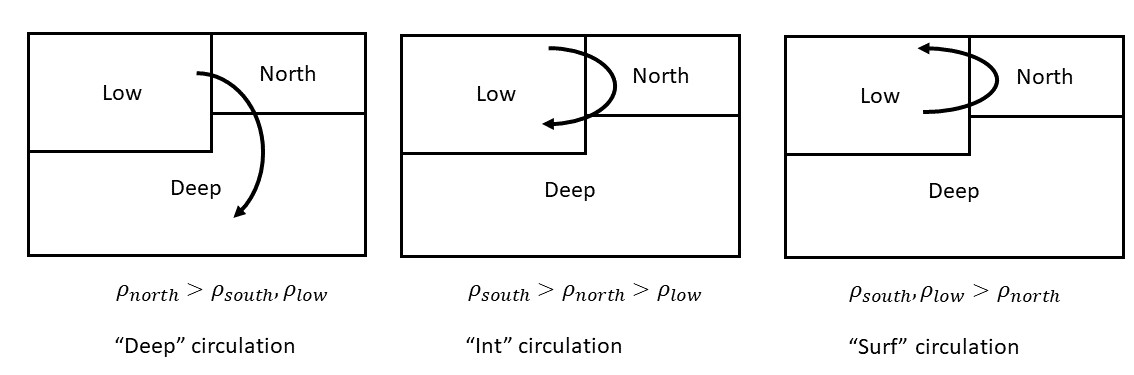}\\
  \caption{Schematic of flows associated with different relationships between density in the southern, low-latitude and northern surface boxes. "Deep" circulation involves formation of water in the northern basin that is denser than both the Southern Ocean and the low-latitude surface boxes and thus is able to connect to the deep. Dynamically it is characterized by a positive overturning circulation as well as lower resistance $\epsilon$ to overturning. "Intermediate" circulation involves formation of water in the northern basin that is lighter than the Southern Ocean water but heavier than the tropics in that basin, similar to what we find in the Pacific today. Dynamically it is characterized by positive values of overturning but with higher resistance $\epsilon$ . "Surface" circulation involves low-latitude water entering the high latitudes and becoming lighter there- similar to what happens in the Arctic or Baltic today. For Northern Basins, this occurs when the density is lighter than both the low-latitude and Southern Ocean boxes. Dynamically it is characterized by negative values of overturning. }
  \label{fig:schematic_DeepIntSurf}
\end{figure}

We note that this formulation then produces multiple possible configurations for the overturning circulation in each basin, outlined schematically in Fig. \ref{fig:schematic_DeepIntSurf}. When the northern basin is denser than either the low latitude or the Southern Ocean box it forms deep water. We refer to this as a "DA" circulation when it is found in the Atlantic and a "DP" circulation when it is found in the Pacific.  When the northern basin has a density between that of the Southern Ocean and that of the low-latitudes (as is the case for the Pacific today), we consider it as forming intermediate water (and refer to it as "IA" and "IP" for the Atlantic and Pacific respectively). When the northern basin becomes sufficiently fresh, as in the "off" state of the \cite{stommel1961thermohaline} model, low-latitude water flows into the basin, gets lighter and is returned to the tropics. We refer to this situation as an "SA" or "SP" circulation for the Atlantic and Pacific respectively. Note that the designation refers to the {\em configuration in density space} of the pathways involved, and not necessarily to the {\em magnitude} of the pathways.  We can then combine these designations to define a taxonomy of the Northern Hemisphere overturning. The modern ocean would then be described as "DA-IP", while an ocean in which the freshwater flux is strong enough to produce freshwater caps in both the North Atlantic and Pacific would be described as a "SA-SP" ocean. The circulation during the last Glacial maximum was arguably an "IA-IP" circulation. This taxonomy will become relevant as we discuss solutions on the model in the Results section.

Following \cite{gnanadesikan1999pycnocline} we allow for diffusive closures of the $M_{eddy}$ and $M_{upw}$ terms
\begin{align}\label{eq:GMflux}
M_{eddy}^{atl,pac} &=\frac{A_{GM} L_x^{atl,pac}}{L_y^s D_{low}^{atl,pac}} , \\
\label{eq:upwelling}
M_{upw}^{atl,pac}&=\frac{K_v A_{low}^{atl,pac}}{D_{low}^{atl,pac}}, 
\end{align}
where $A_{GM}$ is a thickness diffusion coefficient following \citep{gent1990isopycnal}, $K_v$ is a vertical diffusion coefficient,  the $L_x$ terms are the length of the Southern boundary of each basin, and $L_y^s$ is the length scale over which the pycnocline shallows in the south. Note that \cite{levermann2010atlantic} found that in realistic climate models the effective length scale may depend on local density gradients, a process not included in the present study.  

The exchange term is modified from \cite{jones2016interbasin} as 
\begin{equation}\label{eq:exchange}
M_{IB}=\frac{g \left[(\rho_{deep}/\rho_{lpac}-1) {D_{low}^{pac}-
(\rho_{deep}/\rho_{latl}-1) {D_{low}^{atl}}}\right] {\rm min}(D_{low}^{pac},D_{low}^{atl})}{\epsilon_{IB}}
\end{equation}
so that it is proportional to the pressure difference between the two basins integrated over the minimum of the pycnocline depths. Note that this is slightly different than the original \cite{jones2016interbasin} formulation, which turns out to be numerically unstable when the densities are allowed to vary separately in the two basins. Thus having an Indo-Pacific basin that is lighter than the Atlantic  (as is the case in real life) allows for a circulation from the IndoPacific to the Atlantic. We do not consider in this paper the effects of wind stress curl in modulating the pressure gradient at the boundary between the Indian and Pacific Oceans as in \cite{jones2016interbasin}, but this represents a relatively straightforward future extension.  It is worth noting, however, that in order to highlight the importance of these winds, \cite{jones2016interbasin} assumed the density contrast between light and dense waters to be fixed (as was also the case in \cite{gnanadesikan1999pycnocline}). For now the effect of the winds should be thought of as being implicitly included in $\epsilon_{IB}$.

In this paper we also allow the temperatures and salinities to be determined by prognostic equations as in the single basin overturning models of \cite{johnson2007reconciling} and \cite{gnanadesikan2018fourbox}. In addition to the mass fluxes already described, the resulting balance equations also allow for terms due to lateral tracer stirring \citep{redi1982isopycnal}, which produces mixing fluxes between low and high latitude boxes of the general form
\begin{equation}\label{eq:mixing}
    M_{XY}=\frac{A_{Redi} D_{XY} L_x^{XY}}{L_y^{XY}} ,
\end{equation}
where $D_{XY}$ is the depth of the pycnocline at the relevant boundary between the boxes, $L_x^{XY}$ is the length of this boundary and $L_y^{XY}$ is the horizontal length scale over which the pycncoline shallows across the boundary. This mixing flux then produces transports of the form $M_{XY} (T_X-T_Y)$ for temperature and $M_{XY} (S_X-S_Y)$ for salinity.  

We also allow for a mixing flux $M_{SD}$ in the Southern Ocean that simulates the impact of bottom water formation/mixing of intermediate water into the deep. This flux is held constant in all the simulations described in this paper. It allows salt added to the Southern Ocean to escape to the deep ocean rather than being necessarily injected into the low-latitude pycnocline.

Finally, heat exchange between each surface box and the overlying atmosphere is handled using a restoring equation of the form
\begin{equation}
\frac{\partial}{\partial t}(D_X*T_X) = D_{mix}/\tau_{rest}*(T_X^{atm}-T_X)
\end{equation}
Note that this means that the high latitude boxes (for which $D_X=D_{mix}$ are more tightly tied to the atmospheric temperature than are the low-latitude boxes. A full list of of all the parameters is given in Table 1.
\begin{table}[t]
\caption{List of key parameters used in the model. Values that are varied from the control are shown in bold with alternatives shown in italics}
\begin{center}
    \begin{tabular}{lcc}
    Parameter  & Description   & Values\\ \hline
    $Area_{ocean}$& Area of the ocean & 3.6$\times$10$^{14}$  m$^2$\\
    $Area_{latl}$& Area of the low latitude Atlantic  & 0.64$\times$10$^{14}$ m$^2$ \\
    $Area_{lpac}$& Area of the low latitude IndoPacific  & 1.98$\times$10$^{14}$ m$^2$\\
    $Area_{S}$& Area of the Southern Ocean & 0.62$\times$10$^{14}$ m$^2$ \\
    $Area_{natl}$& Area of the N.Atl+Arctic  & 0.22$\times$10$^{14}$ m$^2$\\
    $Area_{npac}$& Area of the Southern Ocean & 0.1$\times$10$^{14}$ m$^2$\\
    $L_x^{satl}$ & Length of Southern boundary of low-latitude Atlantic. & 6.25$\times$10$^{6}$ m  \\
    $L_x^{spac}$ & Length of southern boundary of low-latitude IndoPacific. & 18.75$\times$10$^{6}$ m \\
    $L_x^{natl}$ & Length of northern boundary of low-latitude Atlantic & 5$\times$10$^{6}$ m \\
    $L_x^{npac}$ & Length of northern boundary of low-latitude  Pacific (m) & 10$\times$10$^{6}$ m \\
    $L_y^{satl,spac,natl,npac}$ & Length over which pycnocline shallows & 1$\times$ 10$^6$ m \\
    $D_{oc}$& Depth of ocean & 3680 m \\
    $D_{mix}$ &Depth of High Latitude surface layers & 100 m\\
    $K_v$ & Vertical diffusion coefficient & 1 $\times$ 10$^{-5}$ m$^2$s$^{-1}$ \\
    $A_{GM}$ &Thickness diffusion coefficient & 1000 m$^2$s$^{-1}$ \\
    $A_{Redi}$ &Tracer diffusion coefficient & {\bf 1000},{\it 400,2400} m$^2$s$^{-1}$ \\
    $\epsilon_{natl0,npac0}$ & Baseline resistance to overturning & 1.4 $\times$ 10$^{-4}$s$^{-1}$ \\
    $\Delta \rho_{trans}$ & Width of transition over which resistance increases & 0.1 kg m$^{-3}$ \\
    $\epsilon_{IB}$ & Resistance to interbasin flow & 0.7 $\times$ 10$^{-4}$ s$^{-1}$\\
    $F_w^{natl}$ & Baseline freshwater transport from low-latitude to N. Atl. & 0.45 Sv \\
    $F_w^{npac}$ & Baseline freshwater transport from low-latitude to N. Pac. & {\bf 0.34 Sv},{\it 0.6 Sv} \\
    $F_w^{satl}$ & Baseline freshwater transport from low-latitude Atl. to SO & 0.275 Sv \\
    $F_w^{spac}$ & Baseline freshwater transport from low-latitude Pac to SO & 0.825 Sv \\
    $F_w^{IB}$ &Baseline interbasin flux & 0.15 Sv \\
    $M_{SD}$ &Mixing between Southern surface and deep & 15 Sv \\
    $\tau_{rest}$& Restoring time for temperatures & 1 yr \\
    
    \end{tabular}
\end{center}
\end{table}
\subsection{Calibrating the model}

\begin{figure}[t]
  \noindent\includegraphics[width=33pc]{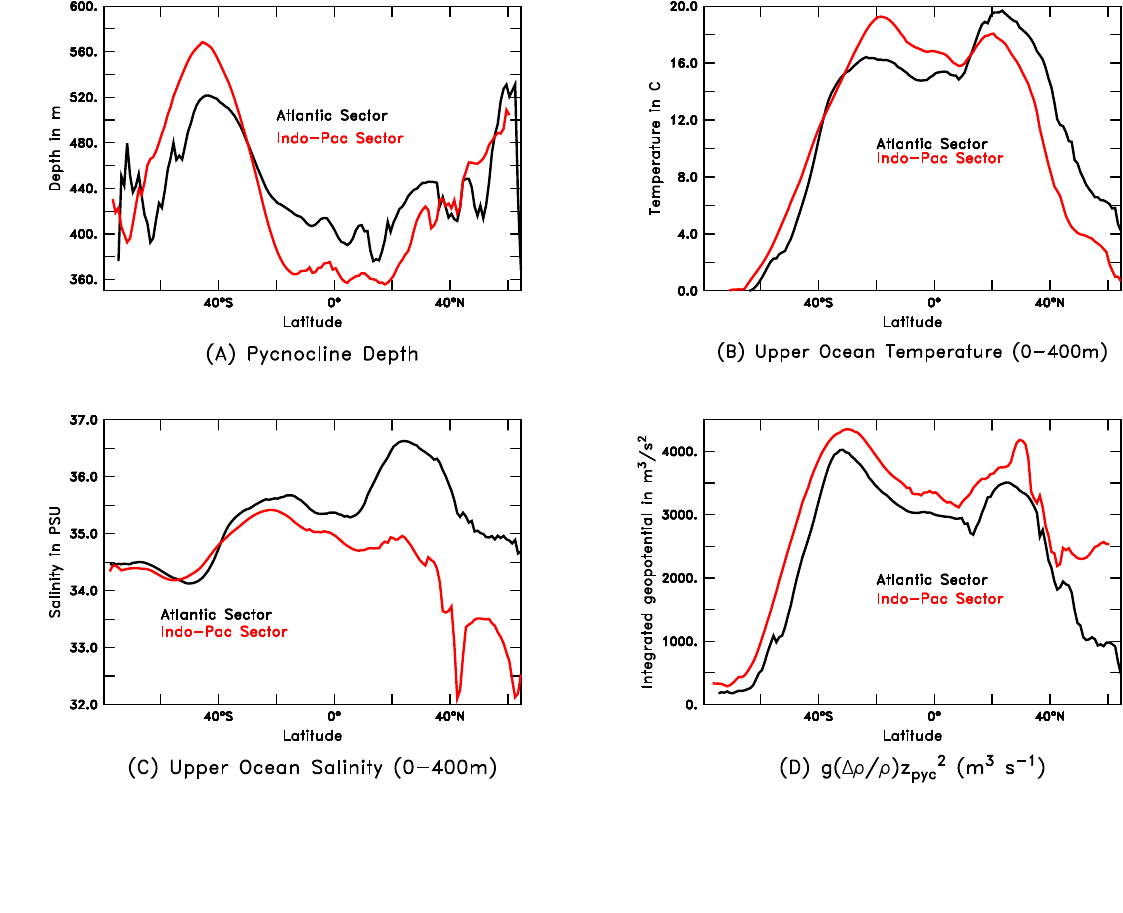}\\
  \caption{Zonally averaged structure of upper ocean hydrography computed from World Ocean Atlas 2013 \citep{WOA2013temp,WOA2013salt}. Black and red lines show values averaged over the Atlantic and IndoPacific respectively.  (a) Pycnocline depth; (b) Upper ocean temperature; (c) Upper-ocean salinity; (d) Upper ocean $g'D_{pyc}^2$.}
  \label{fig:pyc_temp_salt}
\end{figure}

In \citet{gnanadesikan2018fourbox} we calibrated the parameters governing the overturning in the northern hemisphere based on what is known about the rate of overturning and mean watermass properties. As in previous work, we calculate the observed pycnocline depth using data from the World Ocean Atlas 2013 \citep{WOA2013temp,WOA2013salt} as 
\begin{equation}\label{eq:pycnocline}
    D=\frac{\int_{z=-2000}^0 z (\sigma_1(z)-\sigma_1(z = 2000)) dz}{\int_{z=-2000}^0 (\sigma_1(z)-\sigma_1(z=2000)) dz}
\end{equation}
which for an exponential profile gives the e-folding depth. As shown in Fig. \ref{fig:pyc_temp_salt}a the exact value of the horizontal averaged $D_{low}^{atl,pac}$ that we use as a model variable will depend sensitively on the exact bounds of integration. Choosing the range 30$^\circ$S to 45$^\circ$N in the Atlantic and 30$^\circ$S to 40$^\circ$N in the IndoPacific (reflecting the differences in the northward drift of the North Atlantic current vs. the Kuroshio) gives us pycnocline depths of 420m in the Atlantic and 380m in the Pacific.  

Beginning with the Southern Ocean, we have a mean salinity of around 34 PSU and temperature of around 4$^\circ$C for the Antarctic Intermediate Waters. Given that modern deep waters are roughly equally fed from the south and the north, we set $M_s$ to 15 Sv. When we add up all the fluxes entering and leaving the Southern Ocean, we find that we need a freshwater flux of about 1.1 Sv to balance them. We divide the Southern Ocean into an Atlantic sector whose southern boundary comprises 25\% of the total length of the northern edge of the Southern Ocean ($L_x^{satl}=6.25 \times 10^6 m$) and let the IndoPacific comprise the rest of the boundary ($L_x^{spac}=1.875 \times 10^6 m$. We set the freshwater flux to the Southern Ocean in our baseline case to mirror this partitioning, so that $F_w^{satl}=1.1 Sv*0.25=0.275 Sv$ and $F_w^{spac}=1.1Sv*0.75=0.825 Sv$.

We then turn to the Atlantic. The mean temperature and salinity above the pycnocline in the low-latitude Atlantic  are 16.2$^\circ$C and 35.8 PSU respectively in the 2013 World Ocean Atlas. We want our northern box to produce North Atlantic Deep Water, with mean T$\approx$4$^\circ$C and mean S$\approx$ 35 PSU. The integrated geopotential  $g'D^2$ (roughly proportional to APE and shown in Fig. \ref{fig:pyc_temp_salt}d) gives a difference of around 2500 m$^3$ s$^{-1}$. This is roughly consistent with what is seen when we calculate differences latitude by latitude. Given an overturning of 16-20 Sv, this would give us a value of $\epsilon_{natl}=1.235-1.545 \times 10^{-4} s^{-1}$.  For our baseline run we let $\epsilon_{natl}=1.4 \times 10^{-4} s^{-1}$. Taking a baseline value of $A_I=1000$ m$^2$ s$^{-1}$, the mixing flux computed from (\ref{eq:mixing}) is ~2 Sv, so that the freshwater flux required to match the observed salinity difference is 0.45 Sv--quite close to what we find the models of \citet{pradal2014Redi}.  

In the North Pacific, we assume that the there is no deep water formation and so the relevant mass flux is the formation rate of North Pacific Intermediate Water, with a rough transport of about 6 Sv \citep{talley1997north,lumpkin2007global}. Given a low-latitude pycnocline depth of around 380m the low-latitude temperatures and salinities above the pycnocline are 17.2$^\circ$C and 35 PSU respectively. The subpolar waters below the seasonal mixed layer (reflecting North Pacific Intermediate water) have an average temperature of 5.2$^\circ$C and salinity of 33.8 PSU. This then produces a integrated geopotential difference of 1700 m$^2$ s$^{-1}$, resulting in a value of $\epsilon_{npac}=2.8 \times 10^{-4}$s$^{-1}$, about twice that in the North Atlantic. Allowing for an additional mixing flux of ~4 Sv due to the wider basin, this gives us a freshwater flux of 0.34 Sv. Note that this flux is actually smaller than in the North Atlantic despite the greater width of this basin, undermining the assumption made in idealized models that it is the meridional atmospheric flux of freshwater that localizes the overturning to the Atlantic.  Examination of our coupled climate models show the same result when the Arctic is included. 

The difference in the efficiency of overturning between the two basins points to one of the ways in which the $\epsilon$ parameter consolidates the impact of a large number of processes. A key difference between the North Atlantic Deep Water and North Pacific Intermediate Water is that the density of the former  ($\sigma_\theta=27.8$) is much heavier than the latter ($\sigma_\theta=26.7$). The Antarctic Intermediate Water density ($\sigma_\theta=27.0$) lies between the two. This means that the overturning generated by the near-surface APE difference in the Atlantic receives an extra "kick" from Antarctic Intermediate water, implying less resistance to overturning. By contrast, in the Pacific, the Antarctic Intermediate Water slows the overturning, resulting in a larger value of $\epsilon_{npac}$. Rather than focus on directly representing the dynamics of the intermediate water here, we instead represent this effect in our model as a transition between low resistance when the northern basin is denser than the south to a higher resistance when it is lighter. 
\begin{equation}\label{eq:efficiency_transition}
    \epsilon_{natl,npac}=1.4\times 10^{-4} \left[ 1.5+0.5 \tanh \left( \frac{\rho_{natl,npac}-\rho_{s}}{\Delta \rho_{trans}} \right) \right].
\end{equation}

For now, we set $\Delta \rho_{trans}=0.1$ kg m$^{-3}$, which allows for the geometry we want, but recognize that further investigation as in \citet{fuckar2007interhemispheric} is warranted. Finally, the interbasin exchange can be gotten by comparing the integrated geopotential difference relative to the deep water in the two basins. We find that this is around 950 m$^3$ s$^{-2}$. Given an interbasin transport of around 13 Sv \citep{lumpkin2007global}, this implies a resistance $\epsilon_{IB}=7\times 10^{-4} s^{-1}$. This resistance is similar to the Coriolis parameter at the Southern tip of Africa, as would be expected from \citet{jones2016interbasin}. However, as previously noted it may also reflect the effect of wind stress curl in deepening the pycnocline as discussed in this paper. If this were the only process contributing to the contrast between the basins, the interbasin freshwater flux could then be backed out from this transport and the interbasin salinity difference of 0.82 PSU as 0.3 Sv. However the fact that the Atlantic and IndoPacific receive different amounts of relatively fresh Southern Ocean and deep water also contributes to the interbasin difference, and we find a better fit in our model with a baseline value of 0.15 Sv. A full set of the relevant freshwater fluxes is shown in Table 1.

\begin{table}[t]
\caption{Target values (left-hand column) and final circulation for two versions of the box model. Control (center column) has a freshwater flux in the North Pacific $F_w^{npac}$ of 0.34 Sv, smaller than the $F_w^{natl}=0.45 Sv$. Counterfactual case (right-hand column) sets $F_w^{npac}$ to 0.6 Sv, higher than in the North Atlantic, and consistent with what is often found in idealized models }\label{t1}
\begin{center}
\begin{tabular}{lccc}
\hline\hline
Parameter & Observed & Control & Counterfactual\\
\hline
N. Atl. T,S ($^\circ$C,PSU)& 4.0, 35.0 & 4.00,35.06 & 3.98,35.11 \\
LL. Atl. T,S ($^\circ$C,PSU)& 16.2,35.8 & 16.21,35.81  & 16.21,35.86 \\
$D_{low}^{atl}$ (m)& 420 & 429.4 & 427.9  \\
$M_n^{atl}$ (Sv)&16-20& 19.0 & 18.9\\ \hline
N. Pac  T,S ($^\circ$C,PSU) & 5.2,33.8 & 5.19,33.83 & 3.67,31.56 \\
LL. Pac. T,S($^\circ$C,PSU) & 17.2,35.0 & 17.20,34.95  & 17.28,35.02 \\
$D_{low}^{pac}$ (m)&380  & 381.4 & 378,1\\ 
$M_n^{pac}$ (Sv)& 2-8 & 6.4 & -1.7\\ \hline
$M_{IB}$ (Sv) & 11-15  &15.1 & 14.9 \\ \hline
SO T,S ($^\circ$C,PSU) & 4.0,34 & 4.08,34.09 & 4.07,34.10 \\
Deep T,S($^\circ$C,PSU) & 4,34.5 & 4.03,34.5 & 4.02,34.50 \\
\hline
\end{tabular}
\end{center}
\end{table}

Using these baseline parameters we then vary our restoring temperatures in the surface layers to produce a solution that roughly agrees with the target observations.  As shown in Table 2, the temperatures, salinities, densities and transports in this solution do not diverge wildly from modern observations. We also present results from  a counterfactual simulation in which the baseline  $F_w^{npac}$=0.6 Sv.  This configuration results in a DA-SP circulation regime-with a negative overturning in the North Pacific, indicative of warm salty water being converted to light surface waters, and a very fresh and cold surface North Pacific. However, changes in the other basins are relatively small.

\subsection{Numerical Continuation}
In the context of studying the proposed six-box overturning circulation model, continuation algorithms for numerical bifurcation analysis play a crucial role in identifying tipping points. Specifically, these methods help in analyzing the behavior of the overturning circulation as it undergoes a ``hard" bifurcation, such as a saddle-node, i.e. a limit point, or a \emph{subcritical} Hopf bifurcation.
The present discussion focuses on continuation past limit points (saddle-node bifurcations), without aiming to provide a comprehensive guide to all bifurcation scenarios, for which one can refer to a number of published studies \citep{dhooge2008new,doedel2007lecture,doedel2012numerical,fabiani2021numerical}. Consider a parameter-dependent dynamical system, described by a system of \emph{autonomous} ordinary differential equations (ODEs)
\begin{equation}
    \frac{d\bm{y}}{dt}=\bm{f}(\bm{y};\lambda), \qquad f:\mathbb{R}^{n+1}\rightarrow\mathbb{R}^n
    \label{eq:ODEs}
\end{equation}
where $\bm{y}\in\mathbb{R}^n$ is the $n$-dimensional state variable vector, $\lambda \in \mathbb{R}$ is a scalar parameter and the function $\bm{f}$ is time-independent and sufficiently smooth. The goal is to construct a \emph{solution curve} $\Gamma$ for the system of nonlinear algebraic equations:
\begin{equation}
    \Gamma :=\{(\bm{y};\lambda) \in \mathbb{R}^{n+1} \text{ such that } \bm{f}(\bm{y},\lambda)=\bm{0}\},
    \label{eq:equilibria}
\end{equation}
corresponding to the equilibria of the system \eqref{eq:ODEs} for various values of the parameter $\lambda$.
The main concept underlying numerical continuation methods \citep{allgower2012numerical} is to generate a sequence of pairs $(\bm{y}_i,\lambda_i)$, $i = 1, 2, \ldots$ that approximate a specific branch of steady-states, satisfying a chosen tolerance criterion  ($||\bm{f}(\bm{y}_i;\lambda_i)|| \le tol$ for some small $tol>0$) and involves a \emph{predictor-corrector} process.
We start from a known point on the curve, $(\bm{y}_i;\lambda_i) \in \Gamma$, and the tangent vector $\bm{v}_i$ to the curve there, computed through the implicit function theorem.
To compute a new point $(\bm{y}_{i+1};\lambda_{i+1})$ we need two steps: (a) finding an initial guess for $(\bm{y}_{i+1},\lambda_{i+1})$ and (b) iteratively refining the guess to converge towards a point on the curve $\Gamma$ \eqref{eq:equilibria}. We denote the initial guess for $\bm{x}_{i+1}\equiv(\bm{y}_{i+1},\lambda_{i+1})$ as $X_{i+1}^{(0)}$, given by:
\begin{equation}
    X_{i+1}^{(0)}=\bm{x}_i+h\bm{v}_i,
    \label{eq:initial_guess}
\end{equation}
where $h$ is a chosen step size. 
For a small enough $h$ the prediction $X_{i+1}^{(0)}$ is close to the solution curve and can be corrected via e.g. a Newton-like scheme. 
Beyond critical points,
where the Jacobian matrix becomes singular, solution branches can be traced with the aid of numerical bifurcation theory.
For example, solution branches past saddle-node bifurcations (limit points) can be traced by applying the so called pseudo arc-length continuation method. This involves the parametrization of both $\bm{y}$ and $\lambda$ by the arc-length $s$ on the solution curve. The solution is sought in terms of both $\bm{y}(s)$ and $\lambda(s)$ in an iterative manner, by solving until convergence an augmented system, involving eq.~\eqref{eq:equilibria} and the following pseudo arc-length condition:
\begin{equation}
    N(X_{i+1}^{(k)})=(X_{i+1}^{(k)}(s)-X_{i+1}^{(0)})^T \cdot \bm{v}_i=0.
\end{equation}
The tangent vector $v_{i+1}$ to the curve at the new point is then computed; The direction along the curve must be preserved, i.e. $\bm{v}_{i}^T\bm{v}_{i+1}=1$, and $\bm{v}_{i+1}$ must be normalized.
Here, to construct the bifurcation diagrams of the 6-box model, we have employed Cl\_Matcont version 5.4. Cl\_MatCont \citep{dhooge2008new}, a user-friendly Matlab package, that relies on a collection of routines for numerical bifurcation analysis.

\section{Results}

\subsection{Interbasin transport and the sensitivity of the overturning geometry to changes in hydrological amplitude}

\begin{figure}[t]
  \noindent\includegraphics[width=35pc]{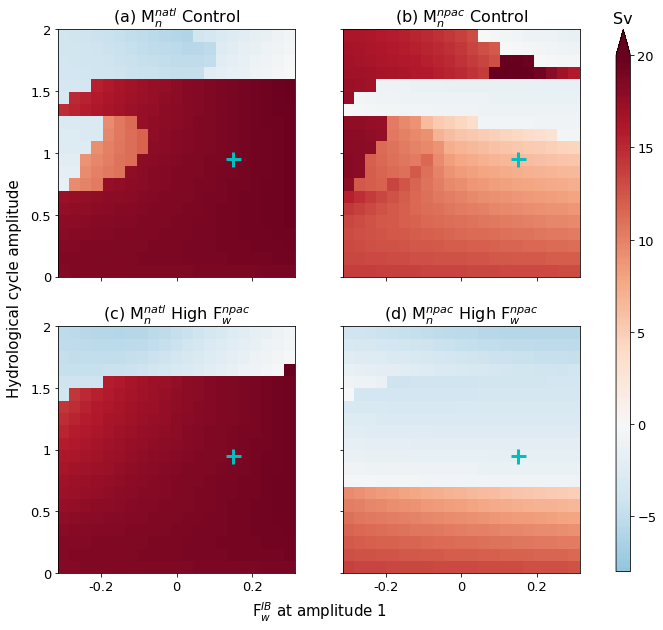}\\
  \caption{Dependence of overturning circulation (Sv) in Atlantic (a and c) and Pacific (b and d) as a function of configuration and amplitude of the hydrological cycle. In each subplot the horizontal axis shows a particular change in the configuration of the hydrological cycle  ($F_w^{IB}$ relative to all the other fluxes). The scale shows the value of $F_w^{IB}$ when the hydrological cycle amplitude is 1. The amplitude of the hydrological cycle is shown along the vertical axis. Plus marks show our estimate of the present-day fluxes. The top row (a and b) shows an experiment where, at the hydrological amplitude of 1, the North Pacific freshwater flux $F_w^{npac}$ is the observationally constrained value of 0.34 Sv (smaller than North Atlantic- so that the poleward freshwater transport acts to make the North Atlantic fresh with respect to the North Pacific). The bottom row (c and d) shows an experiment where $F_w^{npac}$=0.6 Sv at hydrological cycle amplitude of 1, so that the poleward freshwater transport makes the North Atlantic salty with respect to the North Pacific. }
  \label{fig:overturnAtlPac_hydro}
\end{figure}

We begin by examining the interplay between the geometry of the hydrological cycle and the sensitivity of the overturning to instantaneous changes in the amplitude of hydrological cycling. Such changes might be found in an idealized climate model experiment where the greenhouse gas concentrations are suddenly raised or lowered. Starting with our observed initial conditions, we define a set of freshwater flux patterns using the inferred $F_w^{natl,npac,satl,spac}$=0.45, 0.34, 0.275 and 0.825 Sv, respectively, but allow the interbasin transport to vary from -0.3 Sv to +0.3 Sv (horizontal axis). We then take the resulting patterns of freshwater fluxes and scale them up and down, varying from 0.1 to 2 times the "present-day" case (vertical axis). Each model is run for 2000 years. We then repeat this set of experiments using a counterfactual case where the freshwater flux is higher in the Pacific than in the Atlantic ($F_w^{npac}=0.6 Sv$).

Both the Atlantic (Fig. \ref{fig:overturnAtlPac_hydro}a and b) and Pacific overturning (Fig. \ref{fig:overturnAtlPac_hydro}c and d) show a strong sensitivity to the geometry and amplitude of the hydrological cycle. Starting with a baseline present-day configuration in which $F_w^{IB}$=0.15 Sv with hydrological amplitude set equal to 1, we find that we can turn off the overturning in the Atlantic (Fig. 4a) either by increasing the amplitude of the hydrological flux, or by changing the direction of the interbasin atmospheric freshwater transport. The behavior of the Pacific overturning (Fig. \ref{fig:overturnAtlPac_hydro}b) is even more interesting. Starting from our baseline case, we can increase the overturning by decreasing the amplitude of the hydrological cycle (which reduces the contrast in salinity between the Pacific and the Atlantic), reversing the interbasin atmospheric transport so that it goes from the Pacific to the Atlantic (ditto) or {\em increasing} the amplitude of the hydrological cycle. {\em We are thus able reproduce the qualitative behavior whereby the overturning in the Pacific can strengthen in either warmer or colder climates. }

\begin{figure}[t]
  \noindent\includegraphics[width=39pc]{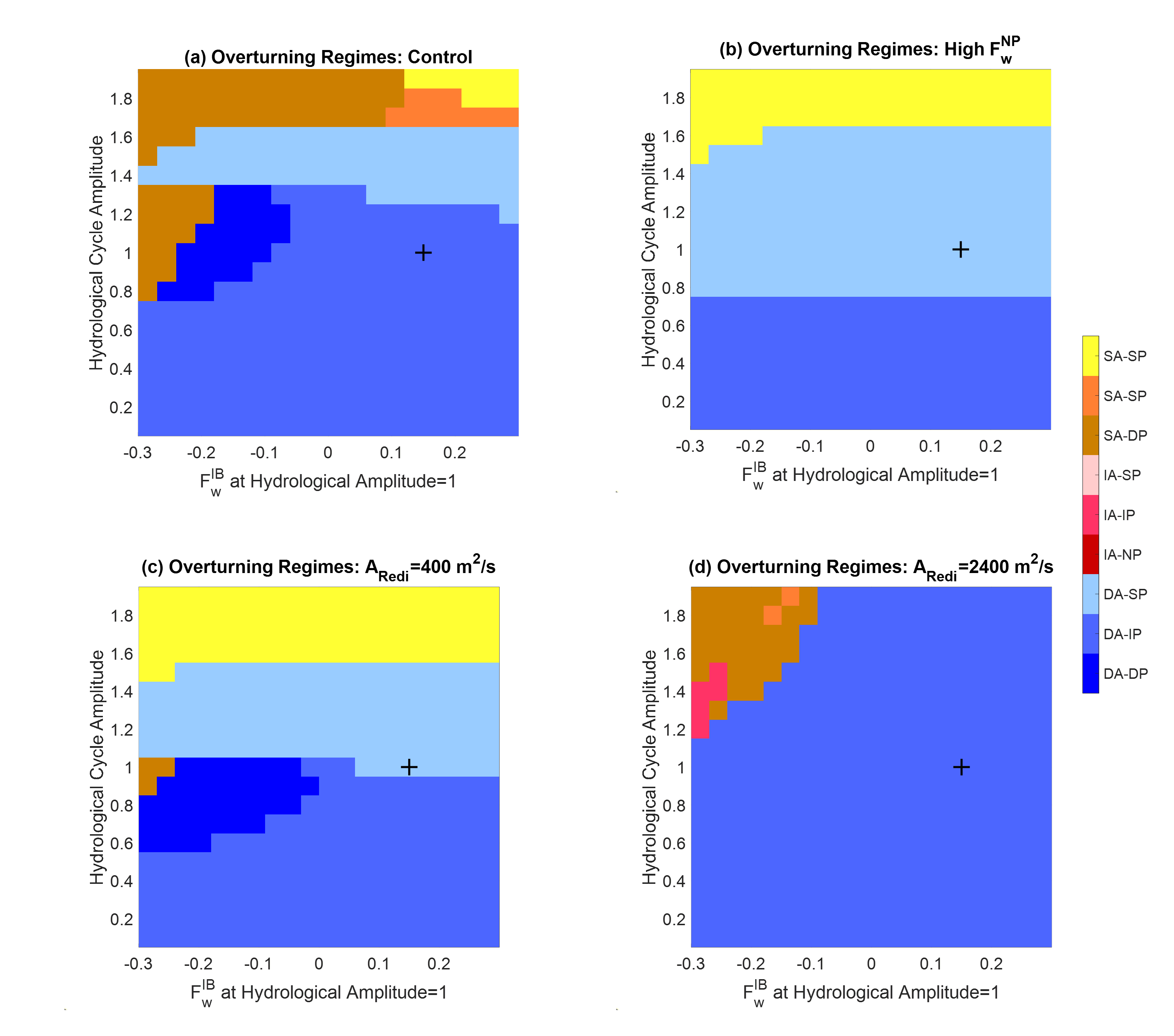}\\
\caption{Classifying global overturning configurations depending on whether Deep, Intermediate or Surface waters are primarily formed in the Atlantic and Pacific. As we move from cooler (blue) to warmer (brown/yellow) colors, Atlantic circulation shallows. As we move from darker shades to lighter ones, Pacific circulation shallows. Axes as in Fig. \ref{fig:overturnAtlPac_hydro}. Plus marks show "present-day" hydrological state. (a) Control simulation ($A_{Redi}=1000 m^2s^{-1}$,$F_w^{npac}=0.34 Sv$). (b) Higher Pacific freshwater flux. $F_w^{npac}=0.6 Sv$. (c) Lower lateral mixing ($A_{Redi}=400 m^2s^{-1}$). (d) Higher lateral mixing ($A_{Redi}=2400 m^2s^{-1}$).}
\label{fig:geometry_plot}
\end{figure}

We can formalize these differences in overturning regimes by constructing a phase diagram of the overturning as a function of hydrological configuration and amplitude. As shown in Fig. \ref{fig:geometry_plot} we color-code the different states, going from cooler to hotter colors as the overturning shallows in the Atlantic and darker to lighter colors as it shallows in the Pacific. In the control case the dominant regime at lower values of hydrological cycling is DA-IP, consistent with the fact that the subpolar N. Pacific is warmer than the Atlantic and Southern Ocean.  A strong enough hydrological cycle with initial conditions similar to today can access either the SA-IP or the SA-SP states (orange colors in upper right of the plot). However, if we reverse the interbasin flux, we can enter a regime  where both basins show deep water formation (dark blue region in Fig.~\ref{fig:geometry_plot}a where the hydrological cycle amplitude is near 1 and there is a moderate flux from the Pacific to the Atlantic). A strong enough reverse flux can produce a SA-DP state (brown).

The counterfactual case shows a much simpler response: overturning shuts down as the amplitude of the  hydrological cycle increases (move from bottom to top)- first in the Pacific (Fig. \ref{fig:overturnAtlPac_hydro}d), then in the Atlantic (Fig. \ref{fig:overturnAtlPac_hydro}c). Changing the interbasin atmospheric water transport $F_w^{IB}$ has relatively little impact on the parameter dependence of overturning over the range shown. However, reversing it so that it dumps freshwater into the Atlantic does result in a Pacific overturning that is slightly more stable to increases in hydrological cycle amplitude and an Atlantic overturning that is slightly less stable. 
The phase diagram for this case (Fig. \ref{fig:geometry_plot}b) shows only three of the six states seen in the Control simulation for the range of parameters covered here- DA-IP at low levels of hydrological cycling, DA-SP at levels comparable to the present day and SA-SP at high levels of cycling.

\subsection{Understanding how the overturning "tips" to the Pacific at high freshwater flux}

\begin{figure}[t]
  \noindent\includegraphics[width=39pc]{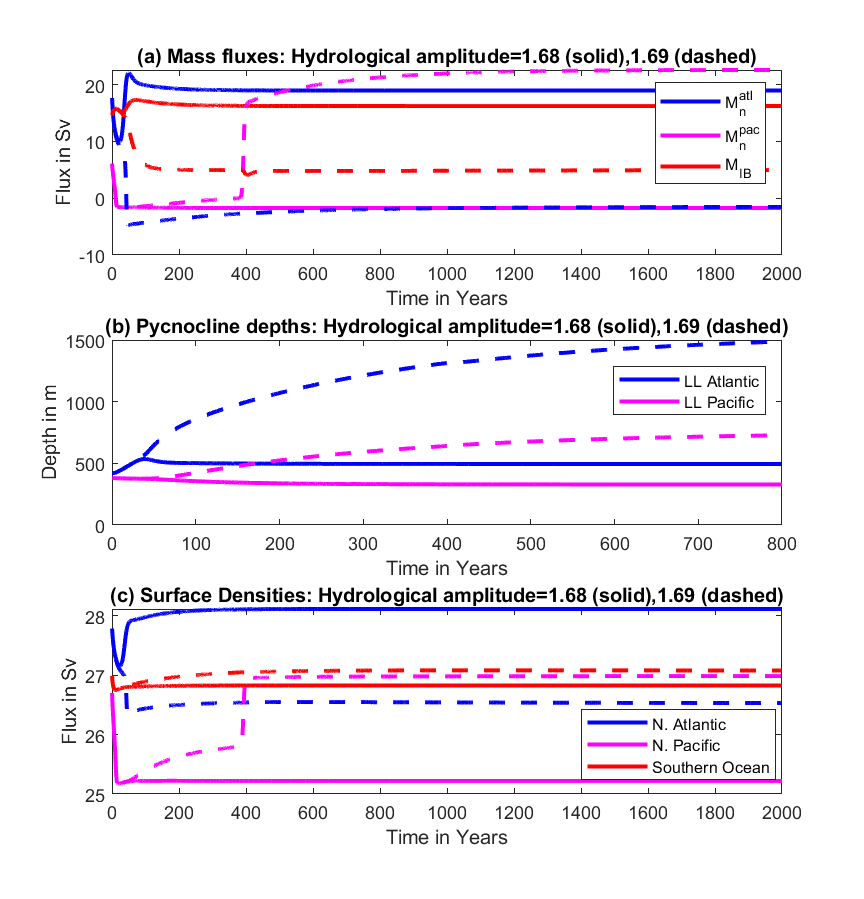}\\
\caption{Evolution of the (a) Large-scale circulation;  (b) Pycnocline depths in Atlantic and Pacific; and (c) densities in two cases near a tipping point. Solid lines show case where a hydrological cycle with the baseline geometry has its amplitude instantaneously increased by a factor of 1.68. Dashed lines show a case where it is increased by a factor of 1.69. }
\label{fig:sixbox_tipping}
\end{figure}

The eventual tipping of the dominant overturning location to the Pacific within our model is sensitive to very small changes in freshwater flux. As shown in Fig. \ref{fig:sixbox_tipping}a, instantaneously increasing the hydrological cycle from our base case by a factor of 1.68 results in a collapse of the North Pacific overturning. There is an initial drop, but then a recovery of the North Atlantic overturning, resulting in a final DA-SP regime. Increasing the scaling factor to 1.69 produces an almost identical initial drop in both overturning circulations, but with the North Atlantic then proceeding all the way to collapse as well, giving us a temporary SA-SP geometry between years 50 and 400. However this means that, as in \citet{johnson2007reconciling}, the transformation of light to dense water in the Southern Ocean is no longer balanced in the North. As a result, the pycnocline in both basins deepens (dashed lines in Fig. \ref{fig:sixbox_tipping}b), causing an increase in the diffusive exchange between the high northern and low latitude boxes. Because the Pacific is wider than the Atlantic, the resulting increase in mixing is slightly faster in the Pacific (at year 200 it is 5.7 Sv in the North Pacific and 4.7 Sv in the North Atlantic). Coupled with the lower freshwater flux in the North Pacific box, this means that the salinity difference in the North Pacific falls a little faster than in the Atlantic, and ends up tipping to a new state as the increase in overturning washes out the salinity difference as in \citet{stommel1961thermohaline}.  Note however, that the North subpolar boxes in both basins stay lighter than the Southern Ocean box, and so the new state corresponds to a SA-IP ocean rather than a SA-DP ocean.

\subsection{Sensitivity to subgridscale eddy mixing}

While there are a host of parameters that can affect the overturning circulation, we focus particularly on one, the lateral diffusion coefficient $A_{Redi}$ that governs the horizontal diffusive exchange of heat and salt between the low-latitudes and the high latitudes. Within a recent generation of climate models $A_{Redi}$ was found to vary from less than 400 m$^2$ s$^{-1}$ to 2000 m$^2$ s$^{-1}$ \citep{abernathey2022isopycnal}.   Previous research \citep{pradal2014Redi,bahl2019variations} has shown that  allowing $A_{Redi}$ to vary between 400 and 2400 m$^2$ s$^{-1}$ can change the stability of the overturning and its intensity within a suite of coupled climate models with similar values of $F_w^{npac}$ and $F_w^{natl}$. Increasing $A_{Redi}$ was found to change the basic regime from a DA-IP regime to a DA-DP regime. However, under global warming the DA-DP regime showed a tendency to collapse the overturning in the Pacific first.

\begin{figure}[t]
  \noindent\includegraphics[width=39pc]{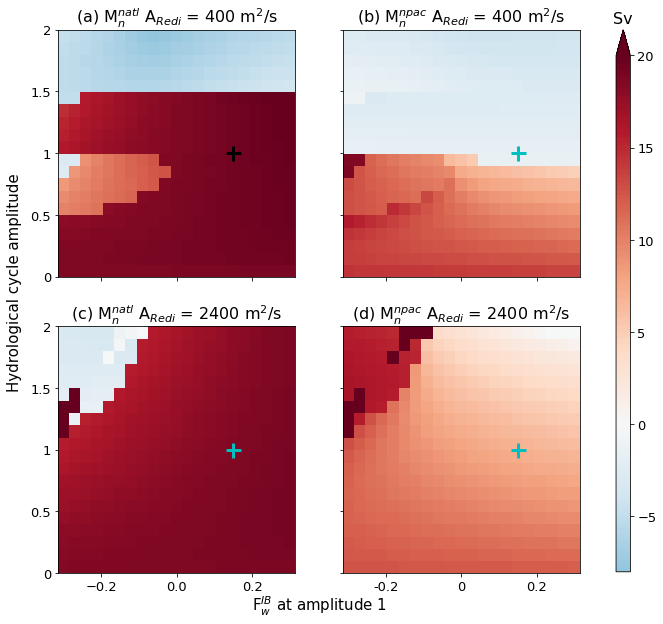}\\
  \caption{Same as Fig. \ref{fig:overturnAtlPac_hydro} except at varying values of  subgridscale lateral eddy mixing parameter $A_{Redi}$. The top row (a and b) shows the Atlantic and Pacific overturning (Sv) respectively, with $A_{Redi}=400$ m$^2$/s, near the lower end of the current value used in climate models. The bottom row (c and d) show the Atlantic and Pacific overturning (Sv) respectively, with $A_{Redi}=2400$ m$^2$/s near the top end of the current value used in climate models.  }
  \label{fig:overturnAtlPac_hydro_Redi}
\end{figure}

Reproducing the sensitivity study of Fig. \ref{fig:overturnAtlPac_hydro}a and b with the control values for $F_w^{npac}$ but with either lower (400 m$^2$s$^{-1}$) or higher (2400 m$^2$s$^{-1}$, bottom row of Fig. \ref{fig:overturnAtlPac_hydro_Redi}) values for $A_{Redi}$ reveals a similarly strong sensitivity to this parameter in the box model.  For lower values of $A_{Redi}$ (top row of Fig. \ref{fig:overturnAtlPac_hydro_Redi}), given modern values of freshwater fluxes, the overturning in the Pacific reverses. The Atlantic overturning with observed fluxes is slightly stronger than with the control $A_{Redi}=1000$ m$^2$ s$^{-1}$ (with a value of 19.3 Sv). However if we change the interbasin freshwater transport (moving to the left from the cross mark) the North Atlantic is more resistant to collapse. It is only a little less stable to increases in the hydrological cycle, with a collapse occurring near an amplitude of 1.58 rather than 1.68.  This can also be seen by comparing the brown areas in Fig. \ref{fig:geometry_plot}a and c. A notable contrast with our control simulation is that turning off the North Atlantic overturning does not result in overturning switching to the Pacific for positive values of interbasin flux (Fig. \ref{fig:geometry_plot}c), because the lower mixing coefficient means that the lateral mixing is not sufficiently strong to degrade the freshwater cap in the Pacific. Only if the interbasin flux reverses do we see the overturning shift to the Pacific.

Increasing the mixing coefficient to 2400 m$^2$ s$^{-1}$ (bottom row of Fig. \ref{fig:overturnAtlPac_hydro_Redi}), on the other hand, produces a somewhat larger (~8 Sv vs 6 Sv in the control) Pacific overturning in the base case which is significantly more stable to changes in both interbasin flux and amplitude of the hydrological cycle. The sensitivity to mixing is much less than in our fully coupled model of \citet{pradal2014Redi}, but we note that in the coupled model the case with $A_{Redi}$=2400 m$^2$s$^{-1}$ predicts a higher density in the North Pacific than in the Southern Ocean. Interestingly, reversing the interbasin freshwater flux with higher mixing can produce an IA-IP state (dark pink, Fig. \ref{fig:geometry_plot}d) which does not appear in the other scenarios.

\subsection{Bifurcation analysis-Baseline case}

\begin{figure}[t]  
  \noindent\includegraphics[trim={3cm 9cm 4cm 9cm},clip,width=0.45 \textwidth]{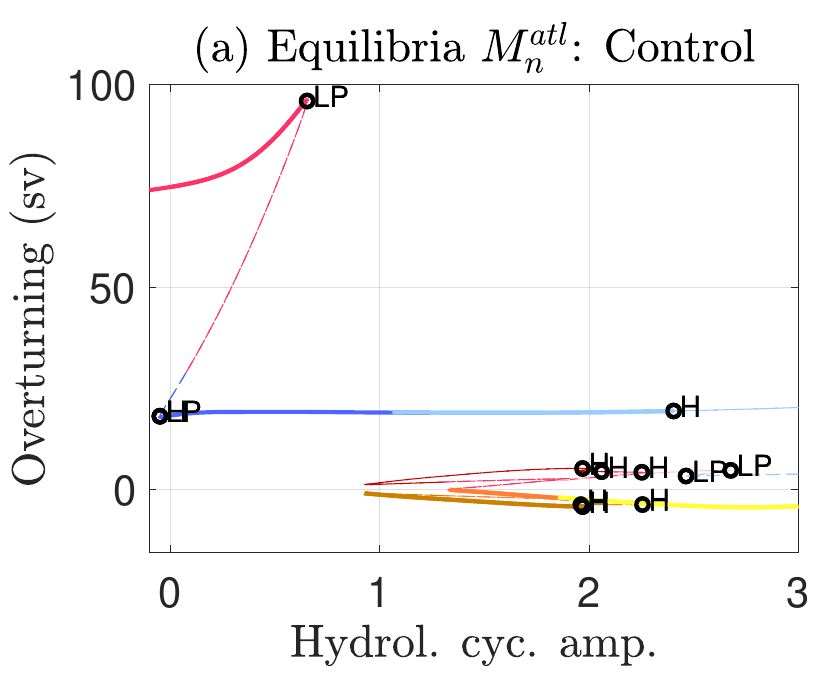}\,
  \noindent\includegraphics[trim={3cm 9cm 4cm 9cm},clip,width=0.45 \textwidth]{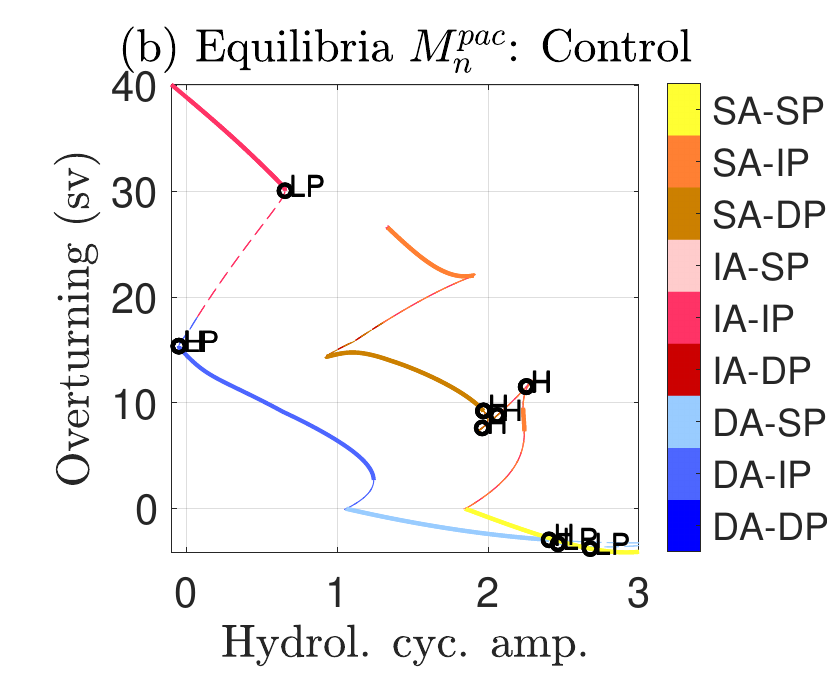}\,
  \noindent\includegraphics[trim={3cm 9cm 3.5cm 9cm},clip,width=0.45 \textwidth]{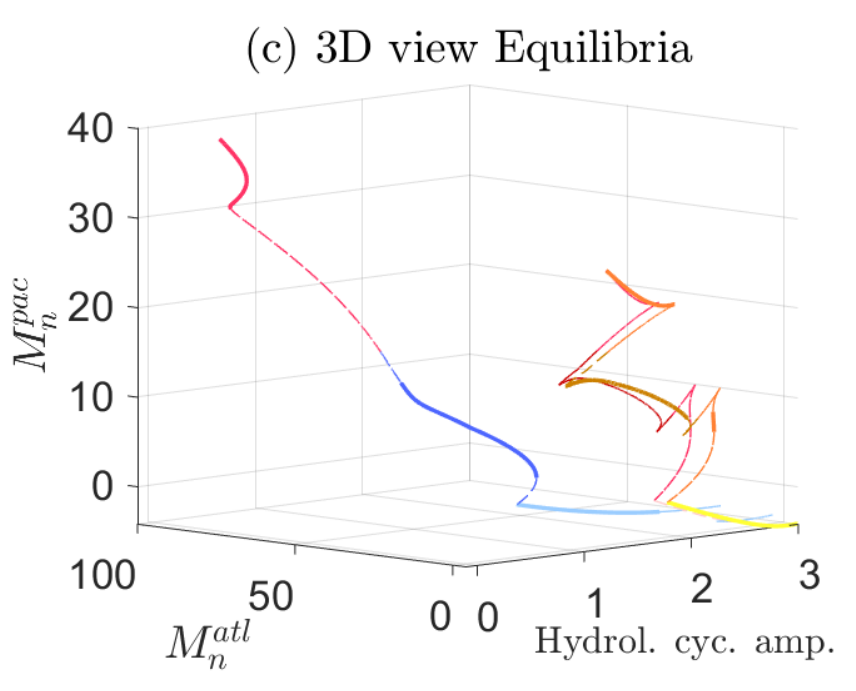}\,
   \noindent\includegraphics[trim={3cm 9cm 3.5cm 9cm},clip,width=0.45 \textwidth]{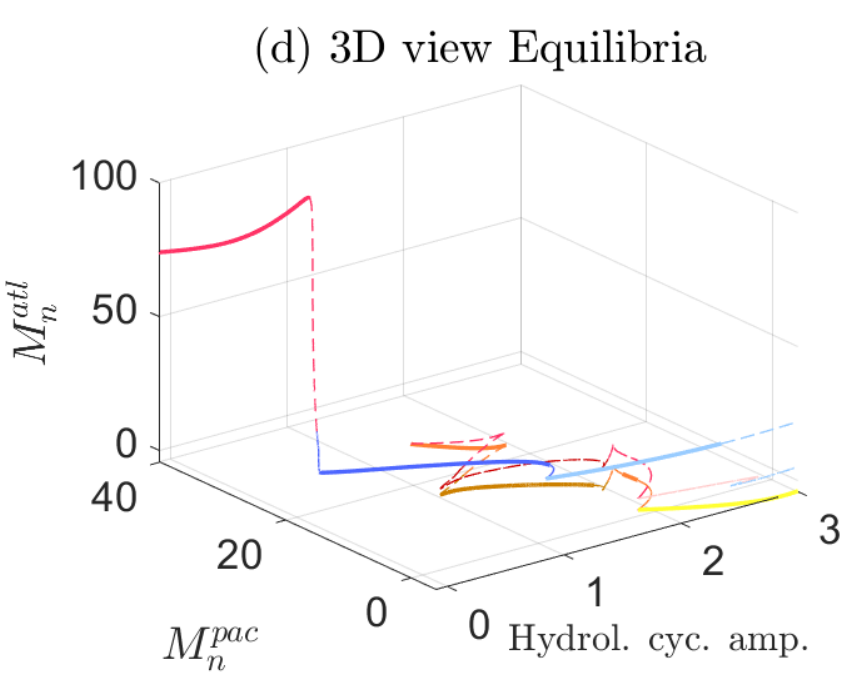}\\ \hspace{1.3cm}
  \caption{Numerical bifurcation diagrams with respect to the amplitude of the hydrological cycle showing steady states of the overturning for the baseline parameter set. All states computed using the matCont code. Colors as in Fig. \ref{fig:geometry_plot}. Limit points are marked with "LP", Hopf bifurcations with "H". Thick solid lines show stable steady states, thinner dashed lines show unstable steady states. (a) North Atlantic overturning $M_n^{atl}$  (b) North Pacific Overturning $M_n^{pac}$ (c,d). Two three-dimensional views of the solution space plotting $M_n{atl}$ and $M_n^{pac}$ against hydrological cycle amplitude.}
  \label{fig:bifurcation_3d}
\end{figure}

\begin{figure}
    \noindent\includegraphics[trim={3cm 9cm 4cm 9cm},clip,width=0.45 \textwidth]{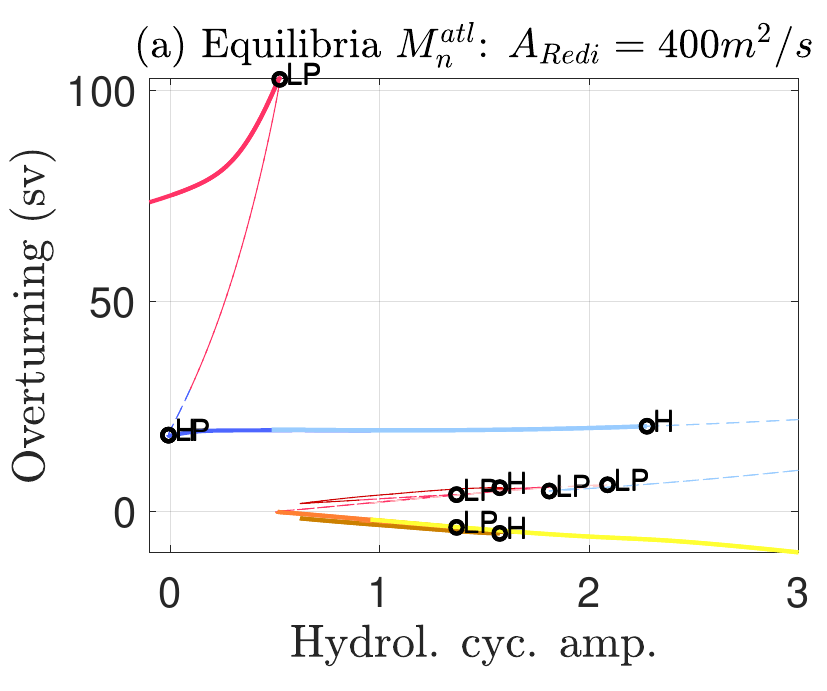}\,
  \noindent\includegraphics[trim={3cm 9cm 4cm 9cm},clip,width=0.45 \textwidth]{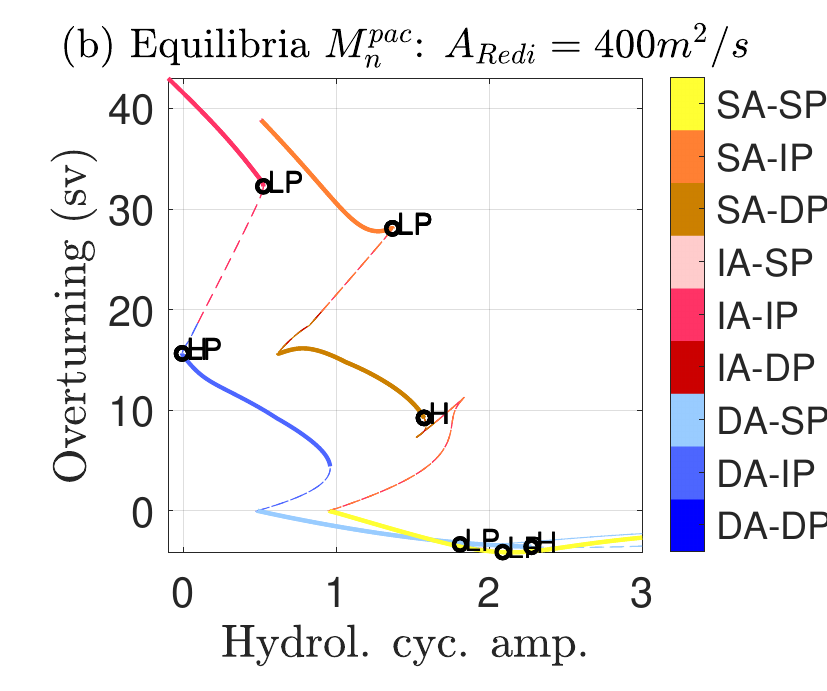}\,
  \noindent\includegraphics[trim={3cm 9cm 4cm 9cm},clip,width=0.45 \textwidth]{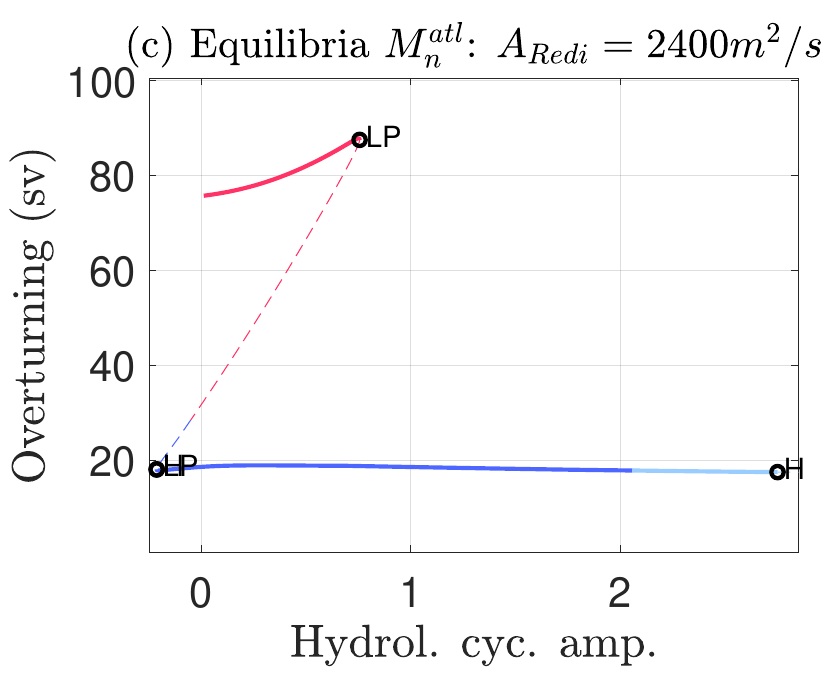}\, \hspace{1.3cm}
  \noindent\includegraphics[trim={3cm 9cm 4cm 9cm},clip,width=0.45 \textwidth]{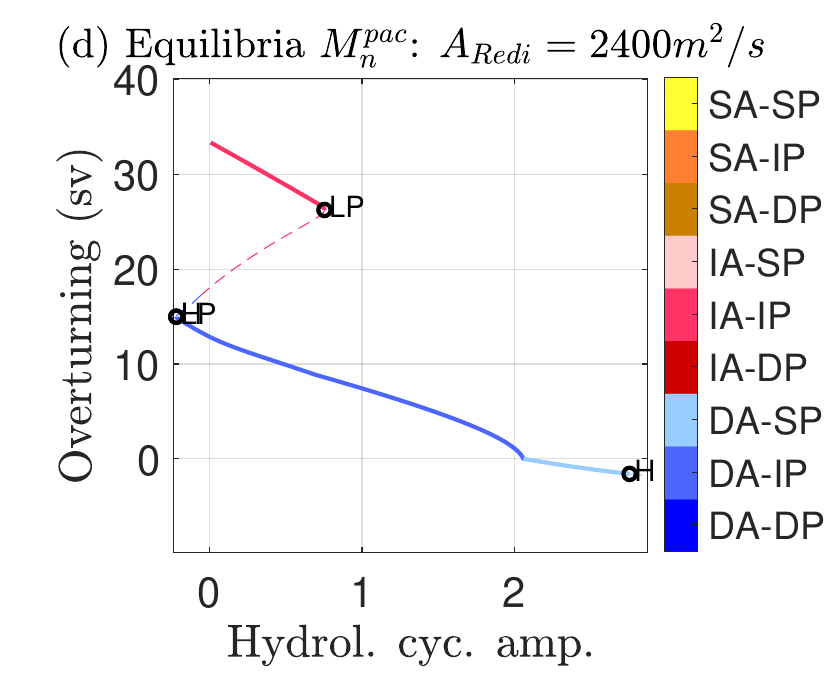}
  \caption{Same as Fig. \ref{fig:bifurcation_3d} a and b but now varying the lateral mixing coefficient  $A_{Redi}$.  (a) Atlantic overturning for low lateral mixing  ($A_{Redi}=400$ m$^2$s$^{-1}$). (b) Same as (a) but for Pacific. (c) Atlantic overturning for high lateral mixing.  ($A_{Redi}=2400$ m$^2$s$^{-1}$). (d) Same as (c) but for Pacific. Colors as in Figs. \ref{fig:geometry_plot}, \ref{fig:bifurcation_3d})}
  \label{fig:bifurcation_Aredi}
\end{figure}

The results presented up to this point assume a particular set of initial conditions (corresponding to the present-day) and an instantaneous change of the hydrological cycle. However, it turns out that if we use numerical continuation methods to allow for quasi-static changes in parameters and to explore a wider range of initial conditions we can find multiple equilibrium states. This is summarized for the baseline parameter set in Fig \ref{fig:bifurcation_3d}, which looks at steady states as functions of the amplitude of the hydrological cycle. In this and the following figure, geometric configurations of the overturning are denoted in the same colors as in Fig. \ref{fig:geometry_plot} with stable branches denoted with thick solid lines and unstable branches denoted with thin dashed ones. The results show a number of surprises.

First, consider what happens if we start from our baseline case  (which we would describe as DA-IP) at a value of 1 for the hydrological cycle amplitude and around 19 Sv for $M_n^{atl}$. As we increase the freshwater forcing (moving to the right along the blue curve in Fig. \ref{fig:bifurcation_3d}a) the overturning in the Atlantic is remarkably steady, while the Pacific overturning collapses.   The collapse of the North Pacific  (Fig. \ref{fig:bifurcation_3d}b) as we increase the freshwater flux looks very much like the classic Stommel fold bifurcation, with a transition to a DA-SP state. Analysis of the eigenvalues of the Jacobian at this point shows that this is, in fact, a limit point bifurcation (marked with LP on the graph). Though it is not visible on this plot, the collapse of the North Pacific overturning actually results in a slight {\em increase} in the North Atlantic overturning. As the freshwater flux continues to increase, there is eventually a Hopf bifurcation (indicated by the H on the figure) at a hydrological cycle amplitude of around 2.3 times the control value, at which point there is a transition to an SA-SP state, with deep pycnoclines in both basins and very fresh northern surface boxes.

These three transitions are not the only possible ones however. For example, there is an additional SA-IP (orange curves) state in which there is a deep pycnocline with overturning only in the North Pacific. This state turns out to be accessible from the SA-SP state by decreasing the hydrological cycle, though interestingly the resulting transition bypasses what looks like an intermediate SA-DP (magenta curves) state when the phase space is examined from this angle. At a hydrological cycle amplitude of around 1.44, the SA-IP state transitions {\em back} to the DA-SP state before retransitioning back to the DA-IP state close to the modern amplitude of the hydrological cycle. Note that there is a symmetry of states around $M_n^{atl}=0$ (Fig. \ref{fig:bifurcation_3d}c and d). These states represent paired solutions in which the high-latitude Atlantic is lighter/heavier than the low-latitude Atlantic, but the absolute value of the density difference and thus the magnitude of the overturning is the same. This accounts for the confusion of colors when looking at only one overturning at a time. 

There is also a branch with very high overturning in both the Atlantic and Pacific at low freshwater flux. This branch turns out to represent an IA-IP case in which there is a very strong flow of warm tropical water to high latitudes in both basins, such that the atmosphere is not able to cool the water enough to make it denser than the Southern Ocean.  For our baseline parameter set, this branch is only stable at hydrological cycle amplitudes lower than the present day.

The dependence of these regimes on hydrological cycle amplitude is sensitive to the parameterization of mixing. For $A_{Redi}=400$ m$^2$s$^{-1}$ (Fig. \ref{fig:bifurcation_Aredi} a,b) the dependence is similar to that at $A_{Redi}$=1000 m$^2$ s$^{-1}$ but all states shift to the left, with SA/SP states permitted at much lower hydrological cycle amplitudes. Interestingly, while the IA-IP geometry collapses at a lower amplitude of hydrological cycle, it allows for higher overturning transports. For the higher mixing case ($A_{Redi}=2400$ m$^2$s$^{-1}$, Fig. \ref{fig:bifurcation_Aredi}c,d) we see the reverse effect: the IA-IP state is weaker at any given hydrological cycle amplitude, but persists to greater hydrological cycle amplitude. At high hydrological amplitudes we no longer see the SA states. Instead, what appears in the model at these parameter ranges is a limit cycle in which there are multicentennial  bursts of DA-SP states which drain the pycnocline interspersed with multicentennial SA-IP states where the pycnocline slowly deepens. Detailed investigation of these states is deferred to a future manuscript.

\subsection{Response to global temperature changes}

Up to this point we have focused on changes in the geometry and amplitude of the hydrological cycle alone, without considering a primary driver of such changes, namely global temperatures. While a full discussion of the sensitivity to global warming is beyond the scope of this manuscript, we present some preliminary exploration of changing temperatures and hydrological cycle together. We consider two cases. In the first, we impose a globally uniform change in the atmospheric restoring temperatures with an associated increase of the hydrological cycle amplitude of 7\% per degree. Poleward water transports in the coupled model simulations reported in \cite{bahl2019variations} are broadly consistent with this sensitivity. In the second set of simulations, we allow for polar amplification in the northern hemisphere high latitudes but reduced warming in the Southern Hemisphere high latitudes. For this set of simulations we use a factor of 1.5 for the high latitude Northern Hemisphere and 0.6 for the high-latitude Southern Hemisphere. These values are typical of the net warming of the atmosphere over the polar oceans in the GFDL ESM2Mc models, and are consistent with the CMIP6 suite of models \citep{hahn2021contributions}. In all cases we started our simulations from the same initial conditions and parameter sets as in the control simulation and integrated for 200 years.

\begin{figure}[t]
  \noindent\includegraphics[width=39pc]{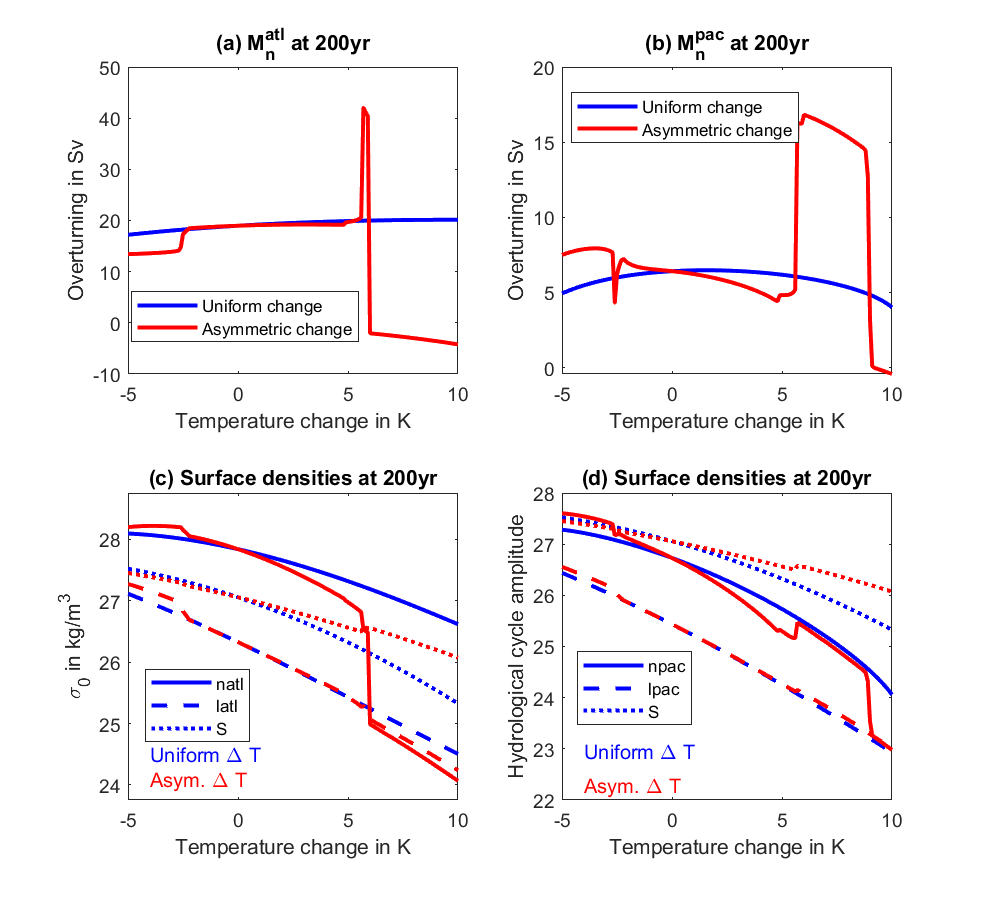}\\
  \caption{Overturning and surface densities given the baseline physical and model initial conditions 200 years after an abrupt change in temperature and associated change in the amplitude of the hydrological cycle. Uniform temperature change is shown in blue, hemispherically asymmetric (1.5 tropical value in north, 0.6 in south) in red. (a) North Atlantic overturning $M_n^{atl}$ in Sv. (b) North Pacific overturning $M_n^{pac}$ in Sv. (c) Densities in North Atlantic (solid), Low Latitude Atlantic (long dashed) and Southern Ocean (dotted). (d) Same as (c) but for Pacific. }
  \label{fig:overturn_response_uniform_asymDT}
\end{figure}

As shown by the blue lines in Fig. \ref{fig:overturn_response_uniform_asymDT}a,b the overturning in both basins is surprisingly insensitive to a uniform change in temperature. This is understandable if we consider the response of the surface densities to climate change (Fig. \ref{fig:overturn_response_uniform_asymDT}c,d). Under uniform temperature change (blue lines), all densities show a similar value of change, with the decrease in density gradient between the northern and low latitudes due to increased freshwater flux as temperature increases largely compensated by the increase in the sensitivity of density to temperature. The idea that higher temperatures might compensate higher freshwater fluxes due to non-linearities of the equation of state was previously advanced by \cite{deBoer2007effect} and \cite{schloesser2020atlantic}. 

Under an asymmetric temperature change, however, the overturning shows more complex behavior (red lines, Fig. \ref{fig:overturn_response_uniform_asymDT}). At low temperatures, the North Pacific becomes denser than than the Southern Ocean due to a combination of a.) the Southern Ocean restoring temperature dropping less than the North Pacific and b.) the fact that in our model the restoring temperature is lower in the North Pacific than in the Southern Ocean to begin with. As a result, deep water can form there. The opening of a second deep water pathway "steals" some of the overturning from the Atlantic for cooling below about 2$^\circ$C (note the dip in the red line at low temperatures in Fig \ref{fig:overturn_response_uniform_asymDT}a). As temperatures warm, on the other hand, the North Atlantic sees its density drop faster than the Southern Ocean. At a warming of around 4.7$^\circ$C  this results in a transition to a more IntNA-IntNP circulation, with an associated deepening of the pycnocline (not shown) and for warming just short of 6$^\circ$C  we see a collapse in both basins and a SA-SP regime. We note that we expect these results to be strongly dependent on the degree of asymmetry in warming, the rate of warming and internal parameters--all of which will be explored in future work.

\section{Discussion}

We have developed a framework for understanding the competing roles of the geometric configuration of the atmospheric hydrological cycle, the amplitude of the atmospheric hydrological cycle, and oceanic eddy processes in setting the geometry and magnitude of the ocean circulation. Key lessons that emerge are: 1. Given a freshwater flux to the Arctic+Subpolar North Atlantic that is larger than the freshwater flux to the subpolar North Pacific, the North Pacific overturning can increase as a result of either increases or decreases in hydrological cycle amplitude. This is likely to be a very different result than would be found from using idealized models with strip continents- highlighting a potential deficiency of such models. 2. We can qualitatively explain the sensitivity of the overturning circulation to the lateral eddy mixing $A_{Redi}$, a parameter that has been previously shown to have an important impact on overturning configuration in fully coupled models. 3. The new geometry and formulations allow for a number of interesting transitions across a range of overturning regimes.

We have only begun to scratch the surface of the parametric dependence of this model. We note that, even for our simple model, we have over twenty initial conditions, physical parameters, and boundary conditions-making a comprehensive exploration of the search space challenging. We are currently exploring two approaches to this. One is to use the continuation methods outlined above to search for interesting phenomena. The second is using generalized adversarial networks to trace out the separatrices in state space between different dynamical states. Early results of the second approach are reported in \citet{sleeman2023generalized}.

In constructing our model we have tried to strike a balance between a parsimonious representation of the processes involved and a sufficiently comprehensive inclusion of key processes.  That said, it should be recognized that there are a number of processes that could benefit from a more sophisticated treatment. One obvious shortcoming of our formulation is how we handle the transition as the Southern Ocean becomes lighter/denser than the high latitude boxes.  Explicitly resolving an intermediate water box as in \citet{alkhayuon2019basin} would allow for a better treatment of this, but would add additional parameters that are harder to constrain. This could be attacked by analyzing experiments with full general circulation models in which the densities of NPIW, AAIW and tropical waters in the Pacific as well as Southern Ocean winds are changed separately, similar to the work of \citet{fuckar2007interhemispheric}, but with realistic geometry and fixed freshwater fluxes. 

Additionally, at this point we have treated the deep ocean as a single box and do not resolve the different impacts of Antarctic Intermediate Water and Antarctic Bottom Water. Incorporating more structure into the deep water \citep[for example following ideas of ][]{nikurashin2011theory} would introduce additional time scales of variability, allow us to better resolve the deep ocean circulation but also introduce additional parametric dependencies on deep mixing.  

Finally, in this paper we have ignored the presence of noise in the climate system. Including noise in the model allows for a rich phenomenology of behavior including unsteady oscillations and more complex transition behavior between states. We plan to report on these phenomena in future publications.

%

%

\clearpage
\acknowledgments
This material is based upon work supported by the Defense Advanced Research Projects Agency (DARPA) under Agreement No. HR00112290032.Approved for public release; distribution is unlimited.

%
%
\datastatement
Matlab codes to generate all the results used in this manuscript are provided at 10.5281/zenodo.8126674.

%






%



\bibliographystyle{ametsocV6}
\bibliography{references}

\end{document}